\documentclass[sigconf, noacm]{acmart} 

\setcopyright{none}
\renewcommand\footnotetextcopyrightpermission[1]{}
\AtBeginDocument{%
  }

\usepackage{enumitem}
\usepackage{amsmath}
\usepackage{booktabs}
\usepackage{array}
\usepackage{caption}
\usepackage{subcaption}
\usepackage{siunitx}
\usepackage{multirow}
\usepackage[table]{xcolor}
\usepackage{graphicx}
\usepackage{etoolbox}
\usepackage{tabularx}
\usepackage{array}

\definecolor{groupbg}{RGB}{255,242,229}
\definecolor{oursbg}{RGB}{231,242,255}

\newcommand{\fakeparagraph}[1]{\noindent\textbf{#1.}}

\begin{document}

\title{WeGR: Towards Unified Generative Retrieval via Behavior-aware Tokenization and Preference Alignment}
\author{
Jiajun Han$^*$,
Guoxin Ma$^*$,
Felix$^{*\dagger}$,
Liangsheng Xu$^\#$,
Ruiqi Yu$^\#$,
Yingdi Guo,
Zongbing Wang,
Xinhua Ye,
Chenyan Hua,
Xin Xie,
Pengcheng Yang,
Haijun Wu
}
\thanks{$^*$These authors contributed equally as co-first authors.}
\thanks{$^\#$These authors contributed equally as co-second authors.}
\thanks{$^\dagger$Corresponding author.}


\renewcommand{\shortauthors}{Jiajun Han et al.}

\begin{abstract}

Generative retrieval (GR) offers a promising paradigm for industrial search by unifying candidate generation within an autoregressive framework. 
However, effectively leveraging user behaviors in industrial GR remains challenging, as existing SID tokenization methods struggle to preserve collaborative signals while maintaining effective codebook utilization, while reinforcement learning (RL) aims to prioritize higher-value candidates based on user feedback but suffers from sparse rewards when generated SIDs fail to match positive items.
To address these challenges, we propose WeGR, a unified generative retrieval framework with two complementary components. Specifically, the \textbf{Co}llaborative and \textbf{Co}llapse-free Tokenizer \textbf{(CoCo-VAE)} injects query-to-item and item-to-item collaborative signals into the continuous representations prior to quantization, thereby preserving behavioral information without interfering with codebook optimization. It further couples codebook refresh, which reactivates underutilized codewords in the first quantization layer, with fidelity anti-collapse regularization, which broadens gradient coverage over semantically relevant codewords and penalizes codeword collisions in deeper layers, yielding behavior-aware SIDs with balanced codebook utilization. For preference alignment, we further introduce \textbf{D}ynamic \textbf{H}orizon \textbf{P}olicy \textbf{O}ptimization \textbf{(DHPO)}, which progressively expands both reward evaluation and policy optimization horizons to alleviate reward sparsity, enabling coarse-to-fine preference learning from SID prefixes to complete items without relying on auxiliary reward models. Extensive offline experiments demonstrate the effectiveness of WeGR in improving retrieval and ranking performance. Moreover, a six-week online A/B test on a large-scale industrial e-commerce search platform demonstrates relative improvements of 2.539\% in orders, 1.508\% in click-through rate, and 1.381\% in conversion rate, while reducing online CPU and GPU usage by 25.5\% and 26.1\%, respectively. These results highlight the effectiveness and efficiency of WeGR in large-scale industrial deployment, supporting multiple search scenarios and processing tens of millions of page views daily.
\end{abstract}



\keywords{Search System, Generative Retrieval, Reinforcement Learning}
\maketitle
\vspace{-0.8em}

\section{Introduction}
\label{sec:introduction}

\begin{figure}[t]
  \centering
  \includegraphics[width=\linewidth]{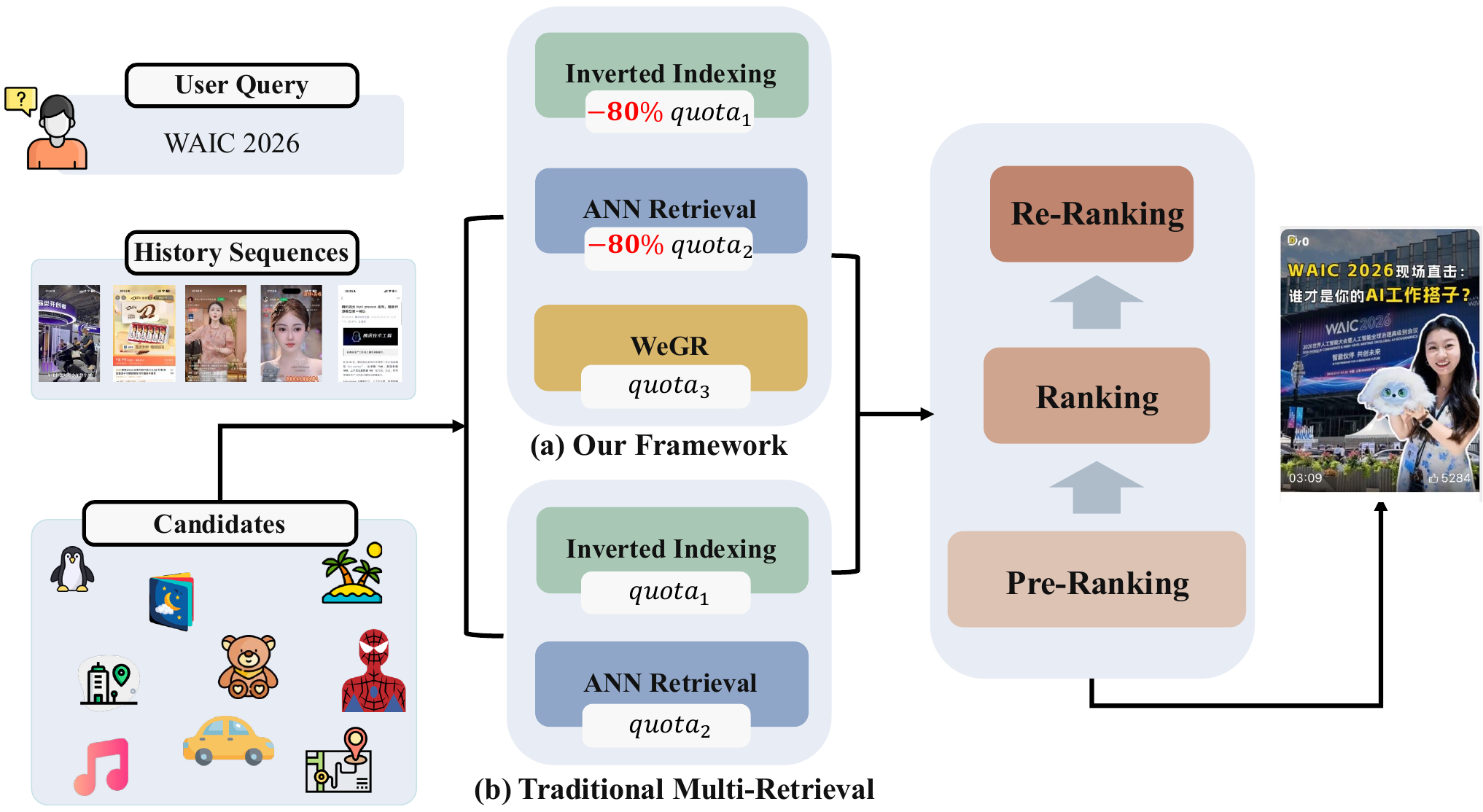}
  \caption{(a) Our proposed End-to-End generative retrieval framework, (b) the traditional multi-stage cascading architecture in E-commerce search.}
  \label{Figure1}
\end{figure}

Modern industrial search systems typically employ multi-stage architectures, where multiple retrieval channels generate candidate items for subsequent ranking stages. While effective, maintaining multiple retrieval pipelines and processing large candidate sets introduce considerable system complexity and computational overhead.
Generative retrieval (GR) emerges as a promising alternative by reformulating candidate generation as an autoregressive process~\cite{chen2025onesearch, 2025onesug, zhang2026onemall, 2025onerec}, where items are represented by sequences of discrete tokens, known as Semantic IDs (SIDs), and retrieved directly conditioned on user queries and contextual information~\cite{2022dsi,nci,seal,2023tiger,xia2026unleash}. As illustrated in Figure~\ref{Figure1}, GR establishes a unified retrieval paradigm capable of simplifying conventional multi-stage pipelines.
However, achieving effective retrieval in large-scale industrial search requires GR to leverage user behaviors at two levels. At the representation level, SID construction should preserve item semantics while incorporating collaborative information from historical co-occurrence patterns. At the generation level, the generative policy should learn from user feedback to distinguish the relative value of candidates and prioritize those that better reflect user preferences.

Existing SID construction methods mainly follow two paradigms. RQ-VAE-based methods enable collaborative supervision during quantization but often suffer from imbalanced codebook utilization and codebook collapse~\cite{2023tiger,wan2026r3}. In contrast, RQ-Kmeans-based methods improve codebook utilization through clustering and balancing strategies~\cite{2025onerec,chen2025onesearch}, yet collaborative signals encoded in upstream item embeddings may be progressively distorted during residual quantization. Consequently, preserving collaborative information while maintaining effective codebook utilization remains a key challenge in SID construction.

Aligning the generative policy with user preferences presents another challenge. Supervised fine-tuning (SFT) maximizes the likelihood of positive SIDs, providing a strong retrieval foundation but failing to explicitly distinguish the relative preference values of generated candidates.
Reinforcement learning (RL) leverages outcome-based rewards, such as clicks and orders, to optimize the generative policy toward prioritizing higher-value candidates~\cite{zhang2026value}.
However, applying Group Relative Policy Optimization (GRPO) with these rewards suffers from severe reward sparsity, as generated SIDs must exactly match positive items to receive feedback~\cite{zhang2026hcgrec}. Existing approaches alleviate this issue through expanded rollouts~\cite{chen2026localsug}, or by introducing estimated relevance and ranking signals to provide denser rewards~\cite{chen2026onesearch,zhang2026onemall}. While the former incurs additional computational overhead, the latter may introduce unreliable supervision, as estimated scores can inherit exposure bias and business-specific heuristics. 
Therefore, the challenge lies in effectively leveraging reliable outcome-based rewards while reducing the difficulty of policy optimization without relying on auxiliary reward models.

To address these challenges, we propose \textbf{WeGR}, a unified generative retrieval framework for industrial search that enhances behavior-aware SID construction and generation-side preference alignment through two complementary components.
Specifically, the \textbf{Collaborative and Collapse-free Tokenizer (CoCo-VAE)} incorporates query-to-item (Q2I) and item-to-item (I2I) collaborative signals into continuous representation learning, while combining codebook refresh with residual fidelity and collision regularization to mitigate codebook collapse and maintain effective codebook utilization. For preference alignment, we introduce \textbf{Dynamic Horizon Policy Optimization (DHPO)}, which employs a coarse-to-fine learning strategy by progressively expanding both reward evaluation and policy optimization horizons. This design alleviates reward sparsity and enables more effective learning from outcome-based rewards, without relying on auxiliary reward models. We further incorporate Multi-Positive Supervised Fine-Tuning (MSFT) to learn from multiple positive items and provide stable supervision when group-relative advantages become uninformative.

Extensive offline experiments and large-scale online A/B tests demonstrate the effectiveness of WeGR in improving retrieval performance and business outcomes while reducing serving costs. 
In a controlled online A/B test, WeGR achieves relative gains of 2.539\% in orders, 1.508\% in click-through rate (CTR), and 1.381\% in conversion rate (CVR), while reducing the number of candidates from conventional retrieval channels by 80\%. 
In terms of serving efficiency, WeGR reduces the number of candidates entering pre-ranking by 38.10\%, lowers CPU and GPU usage by 25.5\% and 26.1\%, respectively, and reduces average serving latency in E-commerce Search by 21.7~ms.
Further analyses across user groups and query frequencies show that WeGR maintains its effectiveness under diverse search conditions.
Beyond these deployment results, we further evaluate the cross-scenario applicability of WeGR by independently examining CoCo-VAE and DHPO in e-commerce live-stream search. CoCo-VAE outperforms RQ-Kmeans, improving HR@6 by 8.28\% and 7.81\% and MRR@6 by 7.07\% and 8.68\% on click and order retrieval, respectively. These gains suggest that its codebook optimization strategy can adapt to different search data distributions while preserving collaborative signals and maintaining effective codebook utilization. Meanwhile, DHPO improves HR@6 by 1.05\% and 3.48\% and MRR@6 by 2.16\% and 3.69\% over SFT, respectively, highlighting the effectiveness of coarse-to-fine preference optimization in alleviating reward sparsity without relying on scenario-specific reward models.
The main contributions of this work are summarized as follows:
\begin{itemize}
    \item We introduce \textbf{CoCo-VAE}, a behavior-aware residual quantization framework that integrates collaborative supervision with codebook refresh, residual-fidelity preservation, and collision regularization to mitigate codebook underutilization without sacrificing SID semantics.
    \item We propose \textbf{DHPO}, a reward-model-free GRPO variant that progressively expands the horizons of reward evaluation and policy optimization along the SID hierarchy, alleviating outcome reward sparsity.
    \item We validate the effectiveness and efficiency of WeGR through extensive offline experiments and large-scale online A/B tests. In online deployment, WeGR achieves significant improvements in business metrics while substantially reducing candidate volume, computational overhead, and serving latency. Additional evaluations in e-commerce live-stream search further demonstrate the cross-scenario applicability of both CoCo-VAE and DHPO, highlighting the practical value of WeGR for industrial search systems.
\end{itemize}

\section{Related Works}

\subsection{Generative Retrieval}

Generative retrieval (GR) formulates retrieval as autoregressive generation, directly predicting target identifiers conditioned on queries or user histories~\cite{2022dsi,nci,seal,p5}. In this paradigm, items are represented by Semantic IDs (SIDs), sequences of discrete tokens that encode item semantics. Existing SID construction methods primarily follow two paradigms: RQ-VAE-based learnable quantization~\cite{2023tiger,2022rqvae} and RQ-Kmeans-based residual clustering~\cite{2025onerec,chen2025onesearch}.

\fakeparagraph{RQ-Kmeans-based methods} These methods construct SIDs through hierarchical clustering of item embeddings and residuals~\cite{luo2024qarm}.  
Representative works include OneRec~\cite{2025onerec}, which improves codebook utilization through balanced clustering, and OneSearch~\cite{chen2025onesearch}, which integrates selective cluster balancing with optimized product quantization~\cite{2014opq}. However, collaborative signals are primarily learned by upstream embedding models rather than being directly incorporated into the quantization process.

\fakeparagraph{RQ-VAE-based methods} These methods jointly learn item representations and discrete codebooks through end-to-end reconstruction and residual quantization, allowing task-specific supervision to guide SID construction. To incorporate collaborative signals, LC-Rec~\cite{2024lcrec} introduces collaborative-semantic alignment to enhance semantic representations with behavioral information, while LETTER~\cite{letter} employs collaborative contrastive learning to align item representations with interaction patterns. ETEGRec~\cite{etegrec} further couples tokenization with downstream recommendation through joint alignment objectives. 
Although these approaches better incorporate behavioral signals into SID construction, RQ-VAE still faces challenges in codebook utilization. Its hard quantization selects only one codeword for each input residual at each layer, leaving rarely selected codewords insufficiently updated and potentially causing codebook collapse. To mitigate this issue, LETTER~\cite{letter} encourages diverse code assignments, DIGER~\cite{diger} introduces recommendation-guided differentiable assignments with Gumbel-noise exploration, and R3-VAE~\cite{wan2026r3} employs reference-vector guidance and scoring-based assignment to stabilize residual quantization.

Despite these advances, existing approaches have yet to fully reconcile collaborative information preservation with effective codebook utilization. CoCo-VAE aims to bridge this gap by integrating behavior-aware representation learning with codebook refresh and complementary regularization.


\subsection{Preference Alignment for GR}


Preference alignment in generative retrieval aims to optimize user preferences and downstream utility beyond generation likelihood. 

Existing approaches can be achieved through either inference-time guidance or training-time optimization. For inference-time guidance, GPR~\cite{zhang2025gpr} employs value-guided trie-based beam search to improve candidate selection without updating the generative policy. 
Training-based approaches instead align the generative policy with user preferences through model optimization. Offline methods leverage historical interactions to learn preference relationships. Reward-weighted SFT~\cite{chidambaram2026robust} and VARG~\cite{varg} incorporate behavioral values into supervised learning, while DPO-based methods~\cite{rafailov2023direct,2024GenR-PO} optimize relative preferences among candidates. RankGR~\cite{fu2026rankgr} and S-DPO~\cite{chen2024softmax} further extend preference learning to listwise comparisons, whereas RAD-DPO~\cite{raddpo} addresses shared-prefix conflicts and noisy feedback. 
In contrast, on-policy reinforcement learning (RL) dynamically samples candidates from the current policy and optimizes generation through reward feedback, typically using PPO~\cite{schulman2017proximal} or GRPO~\cite{shao2024deepseekmath,zhou2023genrrl}. However, learning from outcome-based rewards remains challenging in generative retrieval, as sampled SIDs must exactly match clicked or ordered items to receive positive rewards. Such strict matching often results in sparse rewards and insufficient learning signals~\cite{zhang2026hcgrec}. To alleviate this issue, HCGRec~\cite{zhang2026hcgrec} introduces target-prefix hints to increase the likelihood of rewarding trajectories. OneSearch-V2~\cite{chen2026onesearch} improves token-level credit assignment through position-wise advantages and prefix-dependent gating, while incorporating relevance and CTR estimates into behavioral rewards. Similarly, HRPO~\cite{hrpo} estimates prefix utilities from user-group statistics and decomposes them into residual token credits for fine-grained optimization. Despite improving reward utilization, these approaches may incur additional sampling costs or introduce noise through estimated utility signals.

These limitations motivate us to explore how to effectively utilize outcome-based rewards for preference alignment while reducing optimization difficulty without relying on auxiliary reward models.

\section{Methodology}
\begin{figure*}[t]
    \centering
    \includegraphics[width=\textwidth]{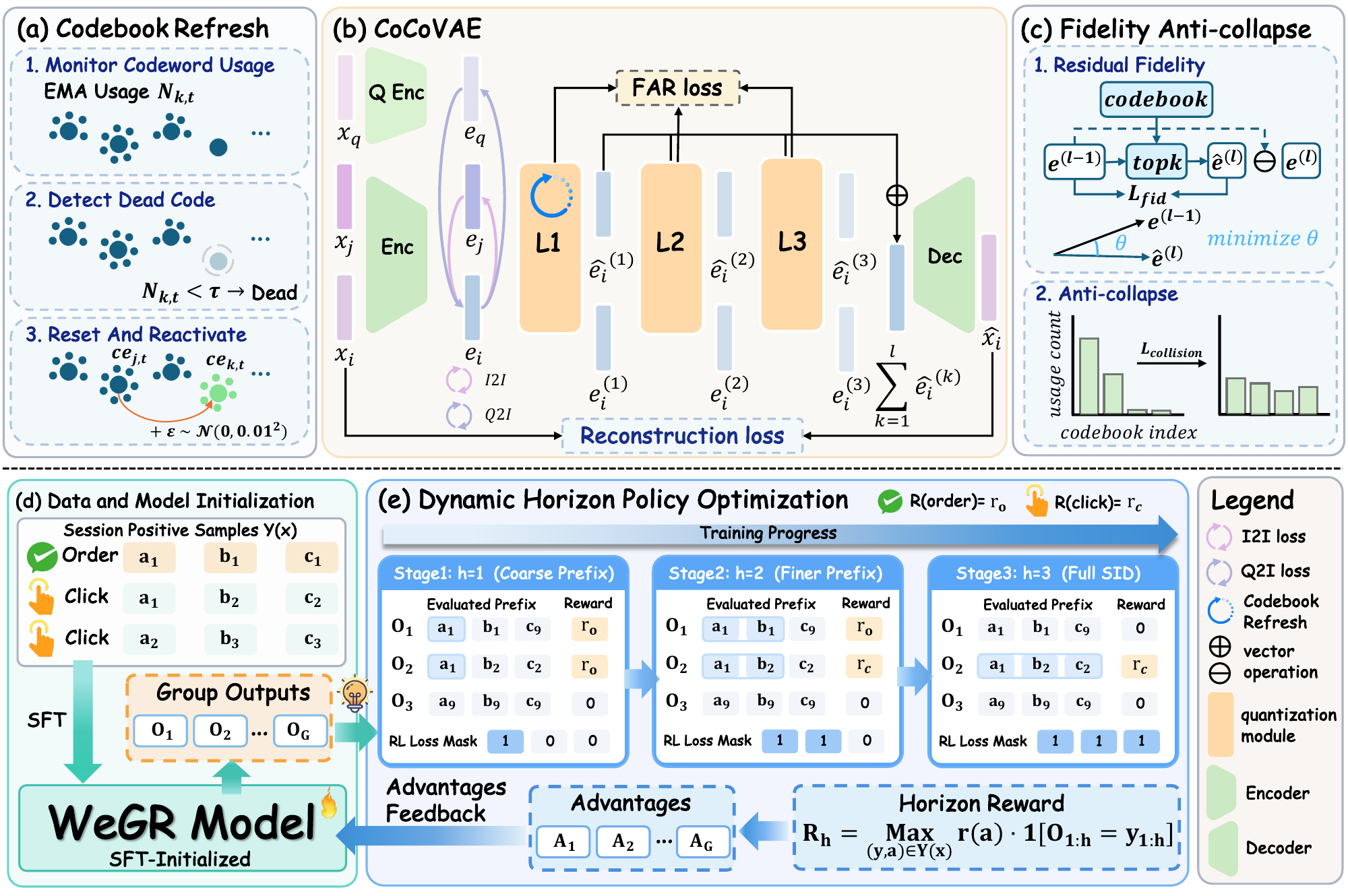}
    \caption{Overview of the proposed WeGR framework. The upper panels (a--c) illustrate CoCo-VAE for behavior-aware SID construction, including (a) codebook refresh to reactivate underutilized codewords, (b) Q2I and I2I collaborative learning to incorporate behavioral co-occurrence signals into item representations, and (c) fidelity and anti-collapse regularization to preserve quantization quality and improve codebook utilization. The lower panels (d--e) present the preference optimization pipeline, covering (d) session-level training data construction and SFT initialization, and (e) DHPO, which progressively expands reward evaluation and policy optimization horizons for coarse-to-fine preference learning.}
    \label{figure_coco_vae}
\end{figure*}

In this section, we propose WeGR, a unified generative retrieval framework for industrial search systems, consisting of CoCo-VAE (\S~\ref{CoCo_VAE}) and DHPO (\S~\ref{sec:dhpo}). The framework is illustrated in Figure~\ref{figure_coco_vae}.

\subsection{Collaborative and Collapse-free Tokenizer}
\label{CoCo_VAE}

To incorporate collaborative signals while ensuring effective codebook utilization, we propose CoCo-VAE, a collaborative-aware residual quantization framework. Specifically, CoCo-VAE incorporates query-to-item (Q2I) and item-to-item (I2I) collaborative signals to guide representation learning, while combining Codebook Refresh with Fidelity Anti-collapse Regularization (FAR) to mitigate codebook collapse during hierarchical quantization.

\subsubsection{Q2I And I2I Collaborative Regularization}

To incorporate collaborative signals into SID construction, we introduce two auxiliary contrastive objectives based on query-to-item (Q2I) and item-to-item (I2I) interactions. Given the input query and item features $\mathbf{x}_q$ and $\mathbf{x}_i$, their respective encoders produce continuous representations $\mathbf{e}_q=f_q(\mathbf{x}_q)$ and $\mathbf{e}_i=f_i(\mathbf{x}_i)$. We construct positive pairs from historical co-occurrence relationships, while using other items within the mini-batch as contrastive candidates.

For Q2I learning, items that co-occur with a query in historical interactions are treated as positive samples. Let $i_q$ denote a positive item associated with query $q$. The Q2I objective encourages the query representation $\mathbf{e}_q$ to be closer to its corresponding positive item representation $\mathbf{e}_{i_q}$ than to other candidate items:

\begin{equation}
\label{eq:q2i}
\mathcal{L}_{\mathrm{Q2I}}
=
-\frac{1}{B_Q}
\sum_{q\in\mathcal{Q}^{+}}
\log
\frac{
\exp\left(\operatorname{sim}(\mathbf{e}_q,\mathbf{e}_{i_q})/\beta_{\mathrm{Q2I}}\right)
}{
\sum_{k=1}^{B}
\exp\left(\operatorname{sim}(\mathbf{e}_q,\mathbf{e}_k)/\beta_{\mathrm{Q2I}}\right)
}.
\end{equation}

Similarly, I2I learning captures collaborative relationships between items that frequently co-occur in user interactions. For each anchor item $j$, let $i_j$ denote its associated positive item. The I2I objective encourages collaboratively related items to have similar representations:

\begin{equation}
\label{eq:i2i}
\mathcal{L}_{\mathrm{I2I}}
=
-\frac{1}{B_I}
\sum_{j\in\mathcal{I}^{+}}
\log
\frac{
\exp\left(\operatorname{sim}(\mathbf{e}_j,\mathbf{e}_{i_j})/\beta_{\mathrm{I2I}}\right)
}{
\sum_{\substack{k=1\\k\neq j}}^{B}
\exp\left(\operatorname{sim}(\mathbf{e}_j,\mathbf{e}_k)/\beta_{\mathrm{I2I}}\right)
}.
\end{equation}

Here, $B$ denotes the number of item representations in a mini-batch, while $\mathcal{Q}^{+}$ and $\mathcal{I}^{+}$ denote the sets of query and item anchors with valid positive pairs, respectively, with $B_Q=|\mathcal{Q}^{+}|$ and $B_I=|\mathcal{I}^{+}|$. The coefficients $\beta_{\mathrm{Q2I}}$ and $\beta_{\mathrm{I2I}}$ control the contrastive temperatures, and $\operatorname{sim}(\cdot,\cdot)$ denotes cosine similarity.

Unlike existing methods~\cite{letter,etegrec} that impose collaborative constraints directly on codebooks, potentially forcing semantically dissimilar items to share the same SID and causing codebook collapse, our approach applies these constraints to continuous representations before quantization, reducing interference with codebook optimization.

\subsubsection{Codebook Refresh}
In residual quantization, some codewords may become ``dead'' due to insufficient usage during training, limiting effective codebook capacity. Incorporating Q2I and I2I collaborative signals may further concentrate item assignments on a few popular codewords, exacerbating codebook imbalance and collapse.
To stabilize the first quantization layer, we introduce Codebook Refresh, a variant of the random-restart strategy~\cite{dhariwal2020jukebox}, as formulated in Eq.~\ref{eq:refresh}. 
We track codeword usage using an exponential moving average (EMA), where $n_{k,t}$ denotes the number of assignments to codeword $k$ in the mini-batch at training step $t$, $N_{k,t}$ represents its accumulated usage statistic maintained through EMA, and $\gamma$ is the decay factor. When $N_{k,t}$ falls below a threshold $\tau$, the codeword is considered ``dead'' and reinitialized using the active codebook embedding with a small Gaussian perturbation. Otherwise, it follows the standard update rule. This process is formulated as:
\begin{equation}
\label{eq:refresh}
\begin{cases}
N_{k,t}=\gamma N_{k,t-1}+(1-\gamma)n_{k,t}, \\
\mathbf{ce}_{k,t}\leftarrow
\mathbb{I}[N_{k,t}\geq\tau]\mathbf{ce}_{k,t}^{\mathrm{update}}
+\mathbb{I}[N_{k,t}<\tau](\mathbf{ce}_{j,t}+\boldsymbol{\epsilon}), \\
\boldsymbol{\epsilon}\sim\mathcal{N}(\mathbf{0},0.01^2\mathbf{I}).
\end{cases}
\end{equation}
where $\mathbf{ce}_{k,t}$ denotes the $k$-th codeword embedding, $\mathbf{ce}_{k,t}^{\mathrm{update}}$ represents its standard update, and $\mathbf{ce}_{j,t}$ is the $j$-th codeword embedding which is more active. This mechanism reactivates underutilized codewords and mitigates persistent codebook collapse. We restrict Codebook Refresh to the first quantization layer $L_1$, as frequent re-initialization in deeper layers may disrupt residual quantization.

\subsubsection{Fidelity Anti-collapse Regularization}
Although Codebook Refresh mitigates codebook collapse in the first quantization layer $L_1$, codewords in deeper layers may remain underutilized. To address this issue, we introduce Fidelity Anti-collapse Regularization (FAR) from two complementary perspectives. At the sample level, we enable each sample to update multiple semantically relevant codewords instead of only its nearest one, expanding gradient coverage. At the codeword level, we regularize codeword assignments across samples by penalizing excessive collisions, preventing a large number of samples from concentrating on a few codewords.

\fakeparagraph{Sample-level Residual Fidelity}
In conventional RQ-VAE~\cite{2023tiger}, hard top-1 assignment limits codeword updates to the nearest codeword, leaving other semantically relevant codewords insufficiently trained. 
To address this limitation, we select multiple codewords most similar to each residual embedding and aggregate them using softmax-normalized similarity weights. Specifically, let $T_i^{(l)}$ denote the set of selected codeword indices for sample $i$ at layer $l$. We construct a weighted codeword representation $\hat{\mathbf{e}}_i^{(l)}$ and minimize its cosine distance from the input residual $\mathbf{e}_i^{(l-1)}$:

\begin{equation}
\label{eq:fid}
\begin{cases}
\mathcal{L}_{\mathrm{fid}} =
\frac{1}{LB}\sum_{l=1}^{L}\sum_{i=1}^{B}
\left[1-\cos(\mathbf{e}_i^{(l-1)},\hat{\mathbf{e}}_i^{(l)})\right],\\[6pt]
\hat{\mathbf{e}}_i^{(l)} =
\sum_{k\in T_i^{(l)}}w_{ik}^{(l)}\mathbf{ce}_k^{(l)},\\[6pt]
w_{ik}^{(l)} =
\frac{
\exp\left(s_{ik}^{(l)}/\beta_{\mathrm{fid}}\right)
}{
\sum_{j\in T_i^{(l)}}
\exp\left(s_{ij}^{(l)}/\beta_{\mathrm{fid}}\right)
}.
\end{cases}
\end{equation}

where $s_{ik}^{(l)}=\operatorname{sim}(\mathbf{e}_i^{(l-1)},\mathbf{ce}_k^{(l)})$ denotes the cosine similarity between the input residual and the $k$-th codeword embedding at layer $l$, and $\beta_{\mathrm{fid}}$ controls the aggregation weights. Here, $B$ and $L$ denote the batch size and the number of quantization layers, respectively. By distributing gradient updates across multiple relevant codewords, this objective improves codebook utilization while preserving residual semantics.

\fakeparagraph{Codeword-level Collision Regularization}
While sample-level residual fidelity broadens gradient updates across relevant codewords, it does not explicitly regulate their overall assignment distribution.
We therefore introduce a codeword-level Anti-Collapse Loss to prevent different samples from concentrating on the same codewords, promoting more balanced codebook utilization. Specifically, we compute the probability distribution of each sample $i$ over all codewords at quantization layer $l$, based on the cosine similarities between its residual embedding and the corresponding codewords:
\begin{equation}
\label{eq:pak}
p_{ik}^{(l)} =
\frac{\exp(s_{ik}^{(l)}/\beta_{coll})}
{\sum_{h=1}^{M_l}\exp(s_{ih}^{(l)}/\beta_{coll})},
\end{equation}
where $p_{ik}^{(l)}$ denotes the soft assignment probability of sample $i$ to codeword $k$ at layer $l$, $M_l$ is the codebook size, and $\beta_{coll}$ is the softmax temperature.
For two distinct samples $i$ and $j$, we measure their codeword collision probability as the overlap between their assignment distributions:
\begin{equation}
\label{eq:cab}
C_{ij}^{(l)}
= \langle \mathbf{p}_i^{(l)}, \mathbf{p}_j^{(l)} \rangle
= \sum_{k=1}^{M_l} p_{ik}^{(l)}p_{jk}^{(l)},
\end{equation}
where $\mathbf{p}_i^{(l)}$ and $\mathbf{p}_j^{(l)}$ represent the probability distributions of samples $i$ and $j$ over all $M_l$ codewords at layer $l$. A higher collision probability indicates a greater likelihood that both samples select the same codeword.

We then minimize the average collision probability across distinct sample pairs and all quantization layers:
\begin{equation}
\label{eq:coll}
\mathcal{L}_{\mathrm{collision}}
= \frac{1}{L}\sum_{l=1}^{L}
\frac{1}{B(B-1)}
\sum_{i\neq j} C_{ij}^{(l)},
\end{equation}
By penalizing codeword collisions across samples, this objective prevents excessive assignments to a few popular codewords, encouraging more balanced codebook utilization.

By combining sample-level residual fidelity with codeword-level collision regularization, we formulate the Fidelity Anti-collapse Regularization (FAR) objective as:
\begin{equation}
\label{eq:sra}
\mathcal{L}_{\mathrm{FAR}}
= \lambda_{\mathrm{fid}}\mathcal{L}_{\mathrm{fid}}
+ \lambda_{\mathrm{collision}}\mathcal{L}_{\mathrm{collision}},
\end{equation}
Together, these two objectives improve codebook utilization by expanding gradient supervision and reducing codeword concentration across samples.

\fakeparagraph{Overall Objective}
The overall training objective combines FAR with the reconstruction loss to preserve item semantic information while mitigating codebook collapse:
\begin{equation}
\label{eq:loss_all}
\mathcal{L}_{\mathrm{all}}
= \lambda_{\mathrm{Rec}}\mathcal{L}_{\mathrm{Rec}}
+ \lambda_{\mathrm{FAR}}\mathcal{L}_{\mathrm{FAR}}
+ \lambda_{\mathrm{Q2I}}\mathcal{L}_{\mathrm{Q2I}}
+ \lambda_{\mathrm{I2I}}\mathcal{L}_{\mathrm{I2I}},
\end{equation}
where $\mathcal{L}_{\mathrm{Rec}}$ denotes the reconstruction loss, $\mathcal{L}_{\mathrm{Q2I}}$ and $\mathcal{L}_{\mathrm{I2I}}$ denote the Q2I/I2I InfoNCE loss\cite{oord2018representation}, and the $\lambda$ coefficients balance the contributions of different objectives.

Together with collaborative regularization and Codebook Refresh, FAR enables CoCo-VAE to preserve behavioral information while improving codebook utilization, yielding more informative and stable SIDs for downstream generative retrieval. As shown in Table~\ref{tab:codebook_metrics}, We compare CoCo-VAE with other SID clustering methods under a unified three-level codebook configuration of (16384,128,128). 
The comparison is based on two codebook-level metrics: independent coding rate (ICR) and codebook utilization rate (CUR). For RQ-OPQ\cite{2014opq}, the corresponding configuration is (16384,128,128,4,4). The R3-VAE\cite{wan2026r3} results were obtained by running its official open-source implementation on our dataset using K-means initialization as recommended in the original paper, whereas the results for the other baselines were obtained using our reimplementations based on their respective papers. 
Note that codebook-level metrics are only intermediate indicators, and the effectiveness of SID construction should ultimately be assessed through downstream retrieval performance, as reported in Table~\ref{tab:main_results}. Although RQ-Kmeans + L3 Balance achieves high codebook utilization, its HR remains relatively low, as forced assignments at L3 may group semantically unrelated items under the same SID, compromising the semantic relevance essential for search-oriented generative retrieval.

\begin{table}[h]
\centering
\caption{Comparison of codebook metrics across different SID methods.}
\label{tab:codebook_metrics}
\setlength{\tabcolsep}{6pt}
\renewcommand{\arraystretch}{1.12}
\resizebox{\columnwidth}{!}{%
\begin{tabular}{lccccc}
    \toprule
    \textbf{Methods}
    & $\mathbf{CUR}_{L1}$
    & $\mathbf{CUR}_{L1*L2}$
    & $\mathbf{CUR}_{\mathit{Total}}$
    & $\mathbf{ICR}$
    \\
    \midrule
    RQ--Kmeans
    & 100\% & 95.56\% & 21.33\% & 6.92\%  \\
    RQ--Kmeans + L3 Balance
    & 100\% & 95.56\% & 52.30\% & 24.26\%  \\
    RQ--OPQ
    & 100\% & 95.56\% & 1.33\% & 22.64\%  \\
    RQ--VAE
    & 99.96\% & 68.11\% & 0.89\% & 0.09\%  \\
    R$^3$--VAE
    & 18.80\% & 18.37\% & 9.19\% & 1.92\%  \\
    \midrule
    CoCo--VAE
    & 99.98\% & 94.85\% & 22.17\% & 7.71\%  \\
    \bottomrule
\end{tabular}%
}
\end{table}

\subsection{Dynamic Horizon Policy Optimization}
\label{sec:dhpo}

To alleviate reward sparsity in generative retrieval, we propose Dynamic Horizon Policy Optimization (DHPO), a coarse-to-fine reinforcement learning framework exploiting SID hierarchies. Instead of directly optimizing complete SIDs, DHPO progressively expands both reward evaluation and policy optimization horizons, enabling a gradual transition from prefix-level to item-level preference learning without auxiliary reward models.

\subsubsection{Dynamic-Horizon Curriculum Learning}
\label{sec:dynamic_horizon}

For each search session with query $q$ and user context $x$, we construct a positive set $\mathcal{Y}(x)=\{(\mathbf{y}^{(j)},a^{(j)})\}_{j=1}^{M_x}$ from historical user interactions, where $\mathbf{y}^{(j)}=(y_1^{(j)},\ldots,y_L^{(j)})$ denotes the $L$-token SID of a clicked or ordered item, $a^{(j)}\in\{\mathrm{click},\mathrm{order}\}$ denotes its corresponding user action, and $M_x$ is the number of positive samples in the session. 
Given a generated SID $\mathbf{o}$, we define the item-level reward as

\begin{equation}
\label{eq:item_reward}
R_{\mathrm{item}}(\mathbf{o},x)
=
\max_{(\mathbf{y},a)\in\mathcal{Y}(x)}
r(a)\cdot
\mathbb{I}\left[\mathbf{o}=\mathbf{y}\right],
\end{equation}

where $r(\cdot)$ assigns rewards $r_o$ and $r_c$ to orders and clicks, respectively, with $r_o>r_c>0$.

DHPO divides policy optimization into multiple stages based on the hierarchical structure of SIDs. We introduce a dynamic horizon $h\in\{1,\ldots,L\}$, which determines the number of SID tokens used for reward evaluation and policy optimization at each stage. Training starts with $h=1$ and gradually extends to the complete SID length $L$, enabling a progressive transition from prefix-level preference learning to item-level ranking optimization. The horizon increases by one after every $\eta$ optimization steps, where $\eta$ denotes the number of training steps in each stage. Formally, at optimization step $n$, the horizon is defined as
\begin{equation}
h(n)=\min\left(1+\left\lfloor\frac{n}{\eta}\right\rfloor,L\right),
\label{eq:horizon_schedule}
\end{equation}

At each training stage, we assign behavioral rewards to generated SIDs based on whether their first $h$ tokens match those of clicked or ordered items in the same session:
\begin{equation}
\label{eq:horizon_reward}
R_h(\mathbf{o},x)
=
\max_{(\mathbf{y},a)\in\mathcal{Y}(x)}
r(a)\cdot
\mathbb{I}\left[\mathbf{o}_{1:h}=\mathbf{y}_{1:h}\right].
\end{equation}
When multiple positive items share the same prefix, the highest reward is retained. At shorter horizons, generated SIDs can receive rewards by matching only the initial tokens of positive items, even when their complete sequences differ. As $h$ increases, DHPO progressively refines preference learning from shorter SID prefixes to complete item-level matching. Importantly, DHPO preserves the original outcome-based rewards while adjusting the matching granularity, avoiding the need for auxiliary reward models.


Given a group of sampled SIDs $\{\mathbf{o}_i\}_{i=1}^{G}$, we compute the group-relative advantage using the rewards evaluated at the current horizon:
\begin{equation}
\label{eq:dhpo_advantage}
\hat{A}_i
=
\frac{
R_h(\mathbf{o}_i,x)
-
\operatorname{mean}_{j=1,\ldots,G}
\left(R_h(\mathbf{o}_j,x)\right)
}{
\operatorname{std}_{j=1,\ldots,G}
\left(R_h(\mathbf{o}_j,x)\right)
+\delta
},
\end{equation}
where $\delta$ is a constant for numerical stability. At each horizon, DHPO compares the rewards of sampled SID prefixes and adjusts their relative generation probabilities to favor those with higher rewards. 

Since prefix matching does not guarantee that the complete generated SID corresponds to a positive item, applying the reward to the entire sequence may cause inaccurate supervision for subsequent tokens. DHPO therefore aligns policy optimization with the reward horizon, updating only the first $h$ tokens to ensure accurate credit assignment. The optimization objective is defined as
\begin{equation}
\small
\mathcal{L}_{\mathrm{DHPO}}
=
-\frac{1}{G}
\sum_{i=1}^{G}
\frac{1}{h}
\sum_{t=1}^{h}
\min
\left(
\rho_{i,t}(\theta)\hat{A}_i,\,
\operatorname{clip}
\left(
\rho_{i,t}(\theta),
1-\epsilon,
1+\epsilon
\right)
\hat{A}_i
\right),
\label{eq:dhpo_loss}
\end{equation}
where $\rho_{i,t}(\theta)=\frac{\pi_{\theta}(o_{i,t}\mid x,\mathbf{o}_{i,<t})}{\pi_{\theta_{\mathrm{old}}}(o_{i,t}\mid x,\mathbf{o}_{i,<t})}$ denotes the per-token importance ratio. 
By jointly expanding the reward and optimization horizons, DHPO gradually extends behavioral supervision to deeper SID positions. When $h=L$, the prefix reward recovers the original item-level outcome reward, i.e., $R_L=R_{\mathrm{item}}$, completing the transition from hierarchical preference learning to exact item-level optimization.

\begin{table*}[t]
\centering
\caption{Overall performance comparison on the Click and Order datasets. The best results are highlighted in bold.}
\label{tab:main_results}
\setlength{\tabcolsep}{4pt}
\renewcommand{\arraystretch}{1.15}
\resizebox{\textwidth}{!}{%
\begin{tabular}{l|cccccc|cccccc}
\toprule
\multirow{2}{*}{\textbf{Method}}
& \multicolumn{6}{c|}{\textbf{Click (20k)}}
& \multicolumn{6}{c}{\textbf{Order (20k)}} \\
\cmidrule(lr){2-7}
\cmidrule(l){8-13}
& HR@6 & HR@50 & HR@200 & MRR@6 & MRR@50 & MRR@200
& HR@6 & HR@50 & HR@200 & MRR@6 & MRR@50 & MRR@200 \\
\midrule

\rowcolor{groupbg}
\multicolumn{13}{c}{\textcolor{blue}{\textit{Semantic ID Construction (SFT)}}} \\
\midrule

RQ--Kmeans
& 0.2641 & 0.4120 & 0.4869 & 0.1714 & 0.1815 & 0.1823
& 0.4051 & 0.5890 & 0.6613 & 0.2702 & 0.2833 & 0.2841 \\

RQ--Kmeans + L3 Balance
& 0.2491 & 0.3746 & 0.4455 & 0.1657 & 0.1745 & 0.1753
& 0.3832 & 0.5374 & 0.6137 & 0.2602 & 0.2711 & 0.2719 \\

RQ--OPQ
& 0.2585 & 0.4000 & 0.4510 & 0.1694 & 0.1790 & 0.1796
& 0.3897 & 0.5661 & 0.6188 & 0.2621 & 0.2744 & 0.2751 \\

RQ--VAE
& 0.1865 & 0.2775 & 0.3185 & 0.1216 & 0.1281 & 0.1285
& 0.3470 & 0.4922 & 0.5416 & 0.2313 & 0.2419 & 0.2425 \\

R$^3$--VAE
& 0.2352 & 0.3717 & 0.4507 & 0.1563 & 0.1656 & 0.1665
& 0.3599 & 0.5389 & 0.6213 & 0.2404 & 0.2536 & 0.2538 \\

\rowcolor{oursbg}
\textbf{CoCo--VAE}
& \textbf{0.2759} & \textbf{0.4293} & \textbf{0.5093}
& \textbf{0.1821} & \textbf{0.1925} & \textbf{0.1934}
& \textbf{0.4103} & \textbf{0.5956} & \textbf{0.6732}
& \textbf{0.2776} & \textbf{0.2909} & \textbf{0.2918} \\

\midrule
\rowcolor{groupbg}
\multicolumn{13}{c}{\textcolor{red}{\textit{Reinforcement Learning (RL)}}} \\
\midrule

DPO
& 0.2693 & 0.4107 & 0.4795 & 0.1785 & 0.1882 & 0.1890
& 0.4051 & 0.5741 & 0.6443 & 0.2781 & 0.2899 & 0.2907 \\

Listwise--DPO (OneSearchV1)
& 0.2701 & 0.4200 & 0.4950 & 0.1787 & 0.1891 & 0.1899
& 0.4042 & 0.5845 & 0.6592 & 0.2747 & 0.2874 & 0.2882 \\

LDPO
& 0.2725 & 0.4172 & 0.4928 & 0.1792 & 0.1893 & 0.1901
& 0.4121 & 0.5840 & 0.6627 & 0.2792 & 0.2912 & 0.2921 \\

Vanilla GRPO
& 0.2588 & 0.3934 & 0.4437 & 0.1691 & 0.1786 & 0.1792
& 0.3910 & 0.5576 & 0.6098 & 0.2639 & 0.2758 & 0.2765 \\

TPMA--GRPO (OneSearchV2)
& 0.2750 & 0.4264 & 0.5067 & 0.1803 & 0.1906 & 0.1915
& 0.4133 & 0.5945 & 0.6722 & 0.2837 & 0.2966 & 0.2974 \\

HCGRec
& 0.2710 & 0.4244 & 0.5045 & 0.1775 & 0.1879 & 0.1888
& 0.4081 & 0.5905 & 0.6664 & 0.2705 & 0.2833 & 0.2842 \\

\rowcolor{oursbg}
\textbf{DHPO (Ours)}
& \textbf{0.2810} & \textbf{0.4304} & \textbf{0.5108}
& \textbf{0.1851} & \textbf{0.1953} & \textbf{0.1962}
& \textbf{0.4217} & \textbf{0.5995} & \textbf{0.6763}
& \textbf{0.2878} & \textbf{0.3003} & \textbf{0.3011} \\

\bottomrule
\end{tabular}%
}
\end{table*}

\subsubsection{Multi-Positive Supervised Fine-Tuning}

Although DHPO alleviates sparse rewards through progressive prefix matching, some rollout groups may still exhibit insufficient reward variation, particularly during early training or for sparse user interactions. Moreover, industrial search naturally involves multiple positive items associated with the same user context,reflecting diverse user interests and behavioral preferences.

To address these limitations, we propose Multi-Positive Supervised Fine-Tuning (MSFT) as a complementary regularization objective. Rather than optimizing a single positive item for each training instance, MSFT jointly supervises all observed positive SIDs:
\begin{equation}
\mathcal{L}_{\mathrm{MSFT}}
=
-\frac{1}{M_x}
\sum_{j=1}^{M_x}
\frac{1}{L}
\sum_{t=1}^{L}
\log
\pi_\theta
\left(
y_t^{(j)}
\mid x,\mathbf{y}_{<t}^{(j)}
\right).
\end{equation}

MSFT complements DHPO by supervising multiple positive SIDs associated with each user context, helping the model capture diverse user interests during preference optimization. When sparse rewards yield uninformative group-relative advantages, MSFT continues to provide effective token-level supervision through next-token prediction, helping preserve the retrieval capability established during SFT initialization.

The overall training objective is
\begin{equation}
\mathcal{L}
=
\mathcal{L}_{\mathrm{DHPO}}
+
\lambda_{MSFT}\mathcal{L}_{\mathrm{MSFT}},
\end{equation}
where $\lambda_{MSFT}$ controls the contribution of the MSFT objective.

\section{Experiment}

\subsection{Experimental Settings}

\fakeparagraph{Datasets}
We extract high-quality user interactions from E-commerce search logs collected between June and July 2026 for supervised fine-tuning (SFT) and reinforcement learning (RL). The dataset spans 31 days, with the first 30 for training and the last for testing. All offline evaluations and ablation studies use this dataset. The Click and Order test sets each contain 20,000 search interactions with observed positive items.

\fakeparagraph{Evaluation Metrics}
Following existing generative retrieval studies~\cite{wan2026r3, chen2025onesearch, chen2026onesearch, 2025onerec}, we evaluate recall and ranking performance using Hit Rate (HR) and Mean Reciprocal Rank (MRR). We report results at $K\in\{6,50,200\}$, averaged over all evaluation samples.

\fakeparagraph{Baseline Methods}
We compare WeGR with SID construction and preference optimization baselines. SID construction baselines include RQ-Kmeans-based methods, namely RQ-Kmeans~\cite{2025onerec}, RQ-Kmeans with balancing, and RQ-OPQ~\cite{chen2025onesearch}, as well as RQ-VAE-based methods such as RQ-VAE~\cite{2023tiger, 2025onesug, letter} and R$^3$-VAE~\cite{wan2026r3}. For preference optimization, we compare DPO~\cite{rafailov2023direct}, S-DPO~\cite{chen2024softmax, chen2025onesearch}, and LDPO~\cite{fu2026rankgr, song2024preference} as off-policy RL baselines, and vanilla GRPO~\cite{shao2024deepseekmath}, TPMA-GRPO~\cite{chen2026onesearch}, and HCGRec~\cite{zhang2026hcgrec} as on-policy RL baselines.

\fakeparagraph{Implementation Details}
We adopt GPT-2 Small~\cite{radford2019gpt2} as the backbone for both offline evaluation and online deployment due to its computational efficiency and low inference latency. For CoCo-VAE, we use three codebook layers with sizes of $(16384, 128, 128)$. The loss weights $\lambda_{\mathrm{fid}}$, $\lambda_{\mathrm{collision}}$, $\lambda_{\mathrm{FAR}}$, $\lambda_{\mathrm{Q2I}}$, $\lambda_{\mathrm{I2I}}$, and $\lambda_{\mathrm{Rec}}$ are set to $0.9$, $1.0$, $0.4$, $0.2$, $0.05$ and $1.0$, respectively. The temperature coefficients $\beta_{fid}$ and $\beta_{coll}$ in $\mathcal{L}_{\mathrm{fid}}$ and $\mathcal{L}_{\mathrm{collision}}$ are set to 1.0 and 0.07, respectively. For RL, we assign rewards of $r_o=2$ and $r_c=1$ to orders and clicks, respectively, and set the MSFT weight to $\lambda=0.5$. We use a GRPO group size of $G=8$ and $\eta=300$ optimization steps per DHPO stage, with $\eta$ adjustable to control stage duration. We update the model daily, performing SFT followed by RL.

\subsection{Offline Performance}

Table~\ref{tab:main_results} summarizes the performance of SID construction and preference optimization methods on the Click and Order datasets. We analyze the results from two aspects.

\fakeparagraph{Effectiveness of SID Construction}
As shown in Table~\ref{tab:main_results}, CoCo-VAE consistently outperforms all SID construction baselines across both Click and Order datasets. Compared with RQ-Kmeans, the strongest SFT baseline, CoCo-VAE improves HR@6 by 4.47\% and 1.28\% on Click and Order, respectively, with further gains of 4.60\% and 1.80\% at HR@200. CoCo-VAE also substantially outperforms existing RQ-VAE-based methods, demonstrating its effectiveness across different quantization paradigms. These improvements highlight the importance of jointly preserving collaborative information and maintaining effective codebook utilization. By mitigating codebook collapse while incorporating user interaction signals, CoCo-VAE can better capture behavioral relationships and distinguish different items in the hierarchical SID space, leading to more accurate item representations and improved recall performance.

\fakeparagraph{Effectiveness of Preference Optimization}
All RL methods are initialized from CoCo-VAE. As shown in Table~\ref{tab:main_results}, existing methods often struggle to improve preference ranking while maintaining retrieval capability. 
DPO-based methods underperform the SFT baseline, mainly because shared SID prefixes between chosen and rejected items lead to conflicting gradients and weaken learning signals. Moreover, DPO may suffer from reward hacking, increasing preference margins by lowering rejected-item probabilities rather than increasing those of chosen items.
For on-policy RL, Vanilla GRPO underperforms SFT across all metrics, as sparse outcome rewards leave many training samples without effective learning signals. 
TPMA-GRPO improves credit assignment through prefix-matching rewards that do not directly reflect user preferences, while HCGRec uses prefix hints to recover learning signals but may introduce a training-inference mismatch.
In contrast, DHPO progressively expands both reward evaluation and policy optimization horizons, reducing learning difficulty while providing reward signals at different levels of the SID hierarchy to guide preference optimization.
Compared with TPMA-GRPO, DHPO achieves relative gains of 2.18\% and 2.03\% in HR@6, and 2.66\% and 1.45\% in MRR@6 on Click and Order, respectively, demonstrating more effective preference alignment while preserving retrieval capability.

\subsection{Ablation Study}
In this section, we conduct ablation studies to evaluate the key components of CoCo-VAE and DHPO.

\fakeparagraph{Ablation of CoCo-VAE}
To verify the effectiveness of each component in CoCo-VAE, we conduct ablation studies on the Click and Order datasets, as shown in Table~\ref{tab:wegr_ablation}. We consider five ablation variants: \textit{i) w/o q2i} removes query-to-item collaborative supervision; \textit{ii) w/o i2i} removes item-to-item collaborative supervision; \textit{iii) w/o fid} removes the residual fidelity loss; \textit{iv) w/o coll} removes the anti-collapse regularization; and \textit{v) w/o refresh} removes the codebook refresh mechanism. The results demonstrate the complementary roles of these components. Removing q2i reduces HR@6 from 0.2759 to 0.2650 on Click and from 0.4103 to 0.3933 on Order, highlighting the importance of query-item co-occurrence in SID construction. Similarly, removing i2i reduces HR@6 to 0.2704 and 0.4042 on Click and Order, respectively, indicating the contribution of item-level collaborative relationships. Moreover, the performance degradation caused by removing residual fidelity—with HR@6 decreasing to 0.2437 and 0.3944 on Click and Order, respectively—or anti-collapse regularization—with HR@6 decreasing to 0.2492 and 0.3987—highlights their respective roles in preserving semantic information and promoting effective codebook utilization. Notably, removing the codebook refresh mechanism leads to the largest performance degradation, with HR@6 dropping to 0.2371 on Click and 0.3712 on Order, demonstrating its critical role in maintaining codebook quality and ensuring effective representation learning. Together, these components enable CoCo-VAE to learn informative SID representations for generative retrieval.

\fakeparagraph{Ablation of DHPO}
To evaluate the contributions of Dynamic-Horizon Curriculum Learning (DHCL) and Multi-positive Supervised Fine-Tuning (MSFT), we consider three ablation variants, as shown in Table~\ref{tab:wegr_ablation}. Specifically, \textit{i) w/o DHCL} removes the progressive horizon scheduling while retaining the joint optimization of GRPO and MSFT; \textit{ii) w/o MSFT} removes the multi-positive supervision while preserving dynamic-horizon optimization; and \textit{iii) w/o both} removes both components, reverting to vanilla GRPO.
The results show that removing both components leads to substantial performance degradation, indicating that directly optimizing sparse outcome-based rewards can undermine the retrieval capability established during SFT. 
Retaining MSFT alone largely recovers this degradation, as learning from multiple positive items provides a stronger foundation for subsequent preference optimization. 
DHCL further improves policy learning by progressively expanding reward evaluation and policy optimization horizons, enabling coarse-to-fine preference learning from SID prefixes to complete items while alleviating reward sparsity. 
With both components, DHPO achieves the best overall performance, improving HR@6 by 1.85\% and 2.78\%, and MRR@6 by 1.65\% and 3.67\% on Click and Order, respectively, compared with the SFT baseline.

\begin{table}[h]
\centering
\small
\setlength{\tabcolsep}{6pt}
\caption{Ablation studies of CoCo-VAE and DHPO. The best results in each group are highlighted in bold.}
\label{tab:wegr_ablation}
\resizebox{0.98\columnwidth}{!}{%
    \begin{tabular}{lcccc}
        \toprule
        \multirow{2}{*}{\textbf{Methods}}
        & \multicolumn{2}{c}{\textbf{Click (20k)}}
        & \multicolumn{2}{c}{\textbf{Order (20k)}} \\
        \cmidrule(lr){2-3}
        \cmidrule(lr){4-5}
        & \textbf{HR@6} & \textbf{MRR@6}
        & \textbf{HR@6} & \textbf{MRR@6} \\
        \midrule
        \rowcolor{groupbg}
        \multicolumn{5}{c}{\textcolor{blue}{\textit{Semantic ID Construction}}} \\
        \midrule
        \rowcolor{oursbg}
        \textbf{CoCo-VAE}
        & \textbf{0.2759} & \textbf{0.1821}
        & \textbf{0.4103} & \textbf{0.2776} \\
        \midrule
        \textit{w/o} q2i
        & 0.2650 & 0.1732
        & 0.3933 & 0.2604 \\
        \textit{w/o} i2i
        & 0.2704 & 0.1748
        & 0.4042 & 0.2693 \\
        \textit{w/o} fid
        & 0.2437 & 0.1585
        & 0.3944 & 0.2602 \\
        \textit{w/o} coll
        & 0.2492 & 0.1627
        & 0.3987 & 0.2644 \\
        \textit{w/o} refresh
        & 0.2371 & 0.1540
        & 0.3712 & 0.2454 \\

        \midrule
        \rowcolor{groupbg}
        \multicolumn{5}{c}{\textcolor{red}{\textit{Reinforcement Learning}}} \\
        \midrule
        \rowcolor{oursbg}
        \textbf{DHPO}
        & \textbf{0.2810} & \textbf{0.1851}
        & \textbf{0.4217} & \textbf{0.2878} \\
        \midrule
        \textit{w/o} DHCL
        & 0.2764 & 0.1830
        & 0.4160 & 0.2835 \\
        \textit{w/o} MSFT
        & 0.2692 & 0.1741
        & 0.4067 & 0.2719 \\
        \textit{w/o} both
        & 0.2588 & 0.1691
        & 0.3910 & 0.2639 \\
        \bottomrule
    \end{tabular}%
}
\end{table}

\subsection{Online A/B Testing}
\label{ABTEST}
To evaluate the effectiveness and efficiency of WeGR in a production environment, we conducted controlled online A/B tests against the incumbent search system from June 22 to August 4. All results in Table~\ref{tab:wegr_online_ab} are reported as relative changes over the control group and are statistically significant ($p<0.05$). We evaluate three deployment configurations that differ in the extent to which WeGR replaces candidates from conventional retrieval channels. \textbf{V1.0 (Additive Retrieval)} introduces WeGR as an additional retrieval channel while retaining all candidates from the other retrievers. \textbf{V1.1 (60\% Candidate Removal)} and \textbf{V1.2 (80\% Candidate Removal)} remove 60\% and 80\% of the candidates contributed by the other retrieval channels, respectively. These configurations allow us to examine whether the business gains persist or improve as WeGR progressively replaces conventional retrieval candidates.

\begin{table}[h]
\centering
\small
\caption{Online A/B testing results from June 22 to August 4. Order, CTR, and CVR denote relative improvements over the production baseline, with bold values indicating statistical significance ($p<0.05$); the last column reports the absolute change in average serving latency.}
\label{tab:wegr_online_ab}
\setlength{\tabcolsep}{3pt}
\begin{tabular*}{\linewidth}{@{\extracolsep{\fill}}lcccc@{}}
\toprule
\textbf{Configuration} & \textbf{Order} & \textbf{CTR} & \textbf{CVR} & \textbf{Latency (ms)} \\
\midrule
V1.0 (Additive) & \textbf{+1.579\%} & \textbf{+1.108\%} & \textbf{+0.520\%} & --- \\
V1.1 (60\% Removal) & \textbf{+1.962\%} & \textbf{+0.884\%} & \textbf{+0.950\%} & $-15.6$ \\
V1.2 (80\% Removal) & \textbf{+2.539\%} & \textbf{+1.508\%} & \textbf{+1.381\%} & $-21.7$ \\
\bottomrule
\end{tabular*}
\end{table}

As shown in Table~\ref{tab:wegr_online_ab}, all three deployment configurations yield statistically significant improvements in Order, CTR, and CVR ($p<0.05$). Under the additive configuration, V1.0 increases Order, CTR, and CVR by 1.579\%, 1.108\%, and 0.520\%, respectively. After removing 60\% of the candidates contributed by the other retrievers, V1.1 further increases the Order and CVR gains to 1.962\% and 0.950\%, while maintaining a significant CTR improvement of 0.884\%. When the removal ratio reaches 80\%, V1.2 yields the largest gains across all three metrics, improving Order by 2.539\%, CTR by 1.508\%, and CVR by 1.381\%. These results show that the business gains do not depend on enlarging the candidate set; instead, the overall performance further improves as WeGR replaces a larger proportion of candidates from conventional retrieval channels.

\begin{figure}[h]
  \centering
  \includegraphics[width=\linewidth]{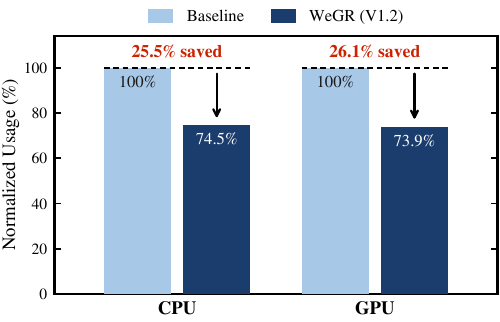}
  \caption{The comparisons of online resource savings for baseline and WeGR.}
  \Description{online resource savings for baseline and WeGR.}
  \label{fig:online_resource_saving}
\end{figure}

\fakeparagraph{Online resource savings}
Beyond the improvements in business metrics, WeGR substantially reduces online serving costs. The key reason is that WeGR replaces, rather than simply supplements, a large fraction of candidates produced by conventional retrieval channels. In V1.2, cutting off 80\% of the candidates from the other retrievers reduces the number of items entering pre-ranking to 61.90\% of the baseline, i.e., 38.10\% fewer items need to be scored. Since pre-ranking  performs item-wise feature processing and model inference over the merged candidate set, this reduction directly decreases its CPU and GPU workloads. As shown in Figure~\ref{fig:online_resource_saving}, the normalized CPU and GPU usages decrease from 100\% to 74.5\% and 73.9\%, corresponding to resource savings of 25.5\% and 26.1\%, respectively. The resource reduction is smaller than the reduction in scored items because the online pipeline still incurs fixed overhead from retrieval, request processing, and other serving components. These results demonstrate that WeGR improves online effectiveness while alleviating computation cost of the downstream cascade.

\fakeparagraph{Online latency reduction}
The last column of Table~\ref{tab:wegr_online_ab} reports the corresponding serving latency. V1.0 leaves the latency unchanged, as it introduces WeGR as an additional retrieval channel without reducing the candidate set entering pre-ranking. Once WeGR starts to replace conventional retrieval candidates, the average latency decreases by 15.6~ms under V1.1 and by 21.7~ms under V1.2. This monotonic trend is primarily attributable to the reduced number of candidates passed to the pre-ranking stage, which lowers the overhead of per-item feature processing and model inference. Together with the business gains, these results demonstrate that WeGR improves effectiveness while reducing serving latency in industrial search systems.

\subsection{Further Analysis}
\label{sec:further_analysis}

In this section, we investigate some questions regarding long-term behavior and user-level benefits of WeGR in industrial deployment.

\begin{figure}[h]
  \centering
  \includegraphics[width=\linewidth]{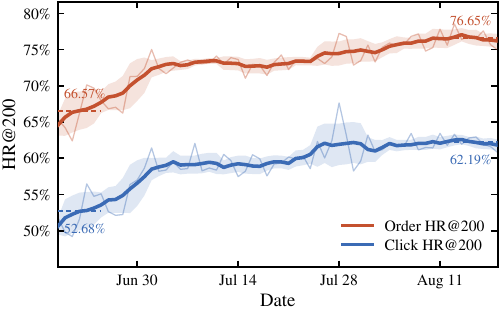}
  \caption{Daily Click and Order HR@200 of WeGR over time.}
  \Description{The daily Click and Order HR@200 of WeGR from June 19 to August 19, 2026.}
  \label{fig:next_day_sid_hr200}
\end{figure}

1) \textbf{How does the performance of WeGR evolve over time?} Industrial search systems operate over a continuously changing item corpus and user interests. It is therefore important to examine whether the retrieval quality of WeGR remains stable as the system evolves. Figure~\ref{fig:next_day_sid_hr200} reports the daily Click and Order HR@200 from June 19 to August 19, 2026, where the bold curves denote the seven-day rolling means and the shaded bands denote the corresponding standard deviations. Despite short-term fluctuations, both metrics exhibit a clear upward trend. Specifically, the average Click HR@200 increases from 52.68\% in the first seven days to 62.19\% in the last seven days, while the corresponding Order HR@200 rises from 66.57\% to 76.65\%, amounting to relative gains of 18.04\% and 15.13\%, respectively. Meanwhile, the shaded bands narrow over time, with the daily standard deviation decreasing from 2.58 to 0.81 percentage points for Click and from 2.64 to 1.09 for Order, indicating that the deployed model becomes not only more accurate but also more stable. Note that these values exceed the offline results reported in Table~\ref{tab:main_results}. This gap mainly stems from the difference in training data scale: the offline evaluation trains on a fixed one-month log, whereas the deployed model is updated periodically and therefore accumulates far more interaction data, which also explains the upward trend observed above. The Order HR@200 remains higher than the Click HR@200 on every day of the observation period, indicating that WeGR effectively retrieves high-value items associated with downstream conversions. 
These results demonstrate that WeGR maintains robust retrieval quality over time and benefits from accumulated interaction signals rather than degrading as the item corpus and user interests evolve.

2) \textbf{Which user groups contribute most to the online gains of WeGR?} We divide users into six groups according to their lifecycle status: active existing users (UG1), reactivated existing users (UG2), lapsed existing users (UG3), active new users (UG4), reactivated new users (UG5), and lapsed new users (UG6). As shown in Table~\ref{tab:user_group_online_gains}, WeGR yields positive Order, CTR, and CVR gains across all six groups, demonstrating that its online improvements are not driven by a single user cohort. The gains are particularly pronounced among reactivated and lapsed new users. UG5 achieves the largest improvements in Order and CTR, reaching 9.855\% and 3.448\%, respectively, while UG6 obtains the highest CVR gain of 7.382\%. UG4 also exhibits substantial increases of 4.823\% in Order and 3.063\% in CVR. In contrast, the gains for active existing users are relatively moderate. This pattern suggests that WeGR is especially effective for users with limited or rapidly changing interaction histories, for whom conventional retrieval methods may suffer from sparse or stale behavioral signals. By jointly exploiting semantic representations and user interactions within a unified generative framework, WeGR can better recover their current search intent and retrieve items with stronger conversion potential.

\begin{table}[h]
  \centering
  \small
  \setlength{\tabcolsep}{4pt}
  \caption{Online relative gains across different user groups. All numbers denote relative improvements over the baseline, and bold values indicate statistical significance ($p < 0.05$).}
  \label{tab:user_group_online_gains}
  \resizebox{0.96\columnwidth}{!}{%
    \begin{tabular}{@{}lcccccc@{}}
      \toprule
      \textbf{Metric} & \textbf{UG1} & \textbf{UG2} & \textbf{UG3} & \textbf{UG4} & \textbf{UG5} & \textbf{UG6} \\
      \midrule
      Order & +1.611\% & \textbf{+5.154\%} & +1.556\% & +4.823\% & +9.855\% & \textbf{+6.143\%} \\
      CTR   & \textbf{+1.125\%} & \textbf{+2.323\%} & \textbf{+2.112\%} & +1.638\% & +3.448\% & +0.524\% \\
      CVR   & +0.399\% & +2.481\% & +0.247\% & +3.063\% & +6.585\% & \textbf{+7.382\%} \\
      \bottomrule
    \end{tabular}%
  }
\end{table}


3) \textbf{On which query-frequency regimes does WeGR work best?} We stratify queries by historical frequency into low ($0$--$3$), medium-low ($4$--$99$), medium ($100$--$999$), and high-frequency ($\geq 1000$) buckets. As shown in Table~\ref{tab:query_frequency_online_gains}, the strongest business-value gains arise at the long-tail and head regions, while the engagement benefit extends to medium-frequency traffic. On long-tail traffic, WeGR significantly improves Order by 3.24\% for low-frequency queries and improves both Order and CVR by 3.86\% and 3.58\%, respectively, for medium-low-frequency queries. These conversion-oriented gains suggest that the learned SID space can transfer collaborative and semantic structure to queries for which direct behavioral evidence is limited. For high-frequency queries, WeGR yields significant gains of 3.51\% in Order and 2.73\% in CTR, indicating that the model also effectively exploits abundant interaction signals for head traffic. Medium-frequency queries exhibit a significant CTR improvement of 0.99\%, despite statistically insignificant changes in Order and CVR, suggesting that increased engagement does not yet translate into downstream conversion within this bucket. Overall, WeGR exhibits complementary effects across the traffic distribution: it primarily strengthens conversion outcomes in the long tail, improves both engagement and orders at the head, and provides a more moderate engagement benefit in the middle. We emphasize that this analysis identifies where the online gains arise; establishing the precise mechanism behind these heterogeneous treatment effects requires further controlled investigation.

\begin{table}[h]
  \centering
  \small
  \caption{Online relative gains across query-frequency buckets. All numbers denote relative improvements over the baseline, and bold values indicate statistical significance ($p < 0.05$).}
  \label{tab:query_frequency_online_gains}
  \begin{tabular*}{\linewidth}{@{\extracolsep{\fill}}lcccc@{}}
    \toprule
    \textbf{Query bucket} & \textbf{Frequency} & \textbf{Order} & \textbf{CVR} & \textbf{CTR} \\
    \midrule
    Low        & $0$--$3$       & \textbf{+3.24\%} & +2.40\%          & +0.95\% \\
    Medium-low & $4$--$99$      & \textbf{+3.86\%} & \textbf{+3.58\%} & +0.91\% \\
    Medium     & $100$--$999$   & $-0.05\%$         & $-0.84\%$         & \textbf{+0.99\%} \\
    High       & $\geq 1000$    & \textbf{+3.51\%} & +1.12\%          & \textbf{+2.73\%} \\
    \bottomrule
  \end{tabular*}
\end{table}

4) \textbf{How does DHPO compare against dense rewards from an auxiliary ranker?}
A straightforward approach to alleviating reward sparsity is to introduce an auxiliary ranker that provides denser rewards for generated candidates. Accordingly, we construct GRPO-Ranker as a baseline, replacing outcome-based rewards with continuous scores predicted by an industrial ranking model. As shown in Table~\ref{tab:dhpo_ranker_comparison}, DHPO consistently outperforms GRPO-Ranker across both click and order retrieval, despite relying solely on outcome-based rewards. This performance gap primarily stems from the candidate distribution mismatch between industrial rankers and GR. Specifically, industrial rankers are trained on subsets filtered by upstream retrieval and pre-ranking stages, whereas GR explores the entire item corpus, including long-tail items with limited historical exposure. Consequently, ranker predictions become less reliable on these under-explored items, where subtle score variations often reflect estimation noise rather than genuine preference differences. Under group-wise advantage normalization in GRPO, such noise is easily magnified into spurious relative advantages, misleading the policy optimization direction. In contrast, DHPO directly leverages authentic outcome-based rewards and dynamically expands both reward evaluation and policy optimization horizons to alleviate reward sparsity. This design not only avoids error propagation from auxiliary reward models, but also eliminates extra ranker inference overhead and reduces dependence on scenario-specific ranking systems.

\begin{table}[h]
\centering
\small
\caption{Comparison of DHPO with ranker-guided GRPO. The best results are highlighted in bold.}
\label{tab:dhpo_ranker_comparison}
\setlength{\tabcolsep}{5pt}
\renewcommand{\arraystretch}{1.12}
\resizebox{0.98\columnwidth}{!}{%
\begin{tabular}{lcccc}
    \toprule
    \multirow{2}{*}{\textbf{Methods}}
    & \multicolumn{2}{c}{\textbf{Click (20k)}}
    & \multicolumn{2}{c}{\textbf{Order (20k)}} \\
    \cmidrule(lr){2-3}
    \cmidrule(lr){4-5}
    & \textbf{HR@6} & \textbf{MRR@6}
    & \textbf{HR@6} & \textbf{MRR@6} \\
    \midrule
    SFT
    & 0.2759 & 0.1821
    & 0.4103 & 0.2776 \\
    GRPO-Ranker
    & 0.2790 & 0.1840
    & 0.4172 & 0.2835 \\
    \midrule
    \rowcolor{oursbg}
    \textbf{DHPO}
    & \textbf{0.2810} & \textbf{0.1851}
    & \textbf{0.4217} & \textbf{0.2878} \\
    \bottomrule
\end{tabular}%
}
\end{table}

5) \textbf{Does WeGR Generalize Across Different Industrial Search Scenarios?}
To evaluate the cross-scenario generalizability of WeGR, we further examine its two core components, CoCo-VAE and DHPO, in e-commerce live-stream search. This setting provides an additional evaluation of both components under different item distributions and user interaction patterns. As shown in Table~\ref{tab:wegr_cross_scenario}, both components consistently outperform their respective baselines across all evaluated metrics.
For SID construction, CoCo-VAE improves HR@6 by 8.28\% and 7.81\%, and MRR@6 by 7.07\% and 8.68\%, over RQ-Kmeans on click and order retrieval, respectively. These gains suggest that its codebook optimization strategy remains effective under different search data distributions while balancing collaborative signal preservation and codebook utilization.
For preference optimization, DHPO improves HR@6 by 1.05\% and 3.48\%, and MRR@6 by 2.16\% and 3.69\%, over SFT on click and order retrieval, respectively. By directly leveraging observed clicks and orders as rewards, DHPO enables coarse-to-fine preference optimization to alleviate reward sparsity without relying on scenario-specific reward models. Its dynamic horizon provides additional learning signals from partial SID matches, allowing preference optimization to progress toward complete item matching.
Overall, the consistent gains of both CoCo-VAE and DHPO demonstrate the effectiveness of WeGR across different industrial search scenarios.

\begin{table}[h]
\centering
\small
\setlength{\tabcolsep}{6pt}
\renewcommand{\arraystretch}{1.1}
\caption{Cross-scenario evaluation of CoCo-VAE and DHPO in live-stream search.}
\label{tab:wegr_cross_scenario}
\resizebox{0.98\columnwidth}{!}{%
    \begin{tabular}{lcccc}
        \toprule
        \multirow{2}{*}{\textbf{Methods}}
        & \multicolumn{2}{c}{\textbf{Click (18k)}}
        & \multicolumn{2}{c}{\textbf{Order (18k)}} \\
        \cmidrule(lr){2-3}
        \cmidrule(lr){4-5}
        & \textbf{HR@6} & \textbf{MRR@6}
        & \textbf{HR@6} & \textbf{MRR@6} \\
        \midrule

        \rowcolor{groupbg}
        \multicolumn{5}{c}{\textcolor{blue}{\textit{Semantic ID Construction}}} \\
        \midrule

        RQ-Kmeans
        & 0.2114 & 0.1428
        & 0.2291 & 0.1670 \\

        \textbf{CoCo-VAE}
        & \textbf{0.2289} & \textbf{0.1529}
        & \textbf{0.2470} & \textbf{0.1815} \\

        \rowcolor{oursbg}
        \textit{Improvement}
        & +8.28\% & +7.07\%
        & +7.81\% & +8.68\% \\

        \midrule
        \rowcolor{groupbg}
        \multicolumn{5}{c}{\textcolor{red}{\textit{Reinforcement Learning}}} \\
        \midrule

        SFT
        & 0.2289 & 0.1529
        & 0.2470 & 0.1815 \\

        \textbf{DHPO}
        & \textbf{0.2313} & \textbf{0.1562}
        & \textbf{0.2556} & \textbf{0.1882} \\

        \rowcolor{oursbg}
        \textit{Improvement}
        & +1.05\% & +2.16\%
        & +3.48\% & +3.69\% \\

        \bottomrule
    \end{tabular}%
}
\end{table}

\section{Conclusion}
In this work, we presented WeGR, a unified generative retrieval framework for industrial search. By harnessing user behaviors from two complementary perspectives, WeGR integrates collaborative signals into SID construction via CoCo-VAE and leverages outcome-based feedback for policy optimization via DHPO, effectively mitigating codebook collapse and reward sparsity. Extensive offline experiments and online A/B tests demonstrate that WeGR significantly improves retrieval quality and business metrics while substantially reducing candidate volume, computational overhead, and serving latency. Additional evaluations in live-stream search further confirm the generalizability of CoCo-VAE and DHPO across distinct data distributions. Overall, these results highlight the potential of WeGR to simplify retrieval pipelines and enable effective, efficient generative retrieval at industrial scale.

\bibliographystyle{ACM-Reference-Format}
\bibliography{sample-base}

@String{Chelsea = "Chelsea" }

@article{2025onesug,
  author       = {Xian Guo and
                  Ben Chen and
                  Siyuan Wang and
                  Ying Yang and
                  Chenyi Lei and
                  et al},
  title        = {OneSug: The Unified End-to-End Generative Framework for E-commerce Query Suggestion},
  year         = {2025},
  url          = {https://doi.org/10.48550/arXiv.2506.06913},
  doi          = {10.48550/arXiv.2506.06913},
  eprinttype    = {arXiv},
  eprint       = {2506.06913}
}

@article{luo2024qarm,
  title={Qarm: Quantitative alignment multi-modal recommendation at kuaishou},
  author={Luo, Xinchen and Cao, Jiangxia and Sun, Tianyu and Yu, Jinkai and Huang, Rui and Yuan, Wei and Lin, Hezheng and Zheng, Yichen and Wang, Shiyao and Hu, Qigen and others},
  journal={arXiv preprint arXiv:2411.11739},
  year={2024}
}

@misc{2022rqvae,
      title={Autoregressive Image Generation using Residual Quantization}, 
      author={Doyup Lee and Chiheon Kim and Saehoon Kim and Minsu Cho and Wook-Shin Han},
      year={2022},
      eprint={2203.01941},
      archivePrefix={arXiv},
      primaryClass={cs.CV},
      url={https://arxiv.org/abs/2203.01941}, 
}

@article{2014opq,
  author       = {Tiezheng Ge and
                  Kaiming He and
                  Qifa Ke and
                  Jian Sun},
  title        = {Optimized Product Quantization},
  journal      = {{IEEE} Trans. Pattern Anal. Mach. Intell.},
  volume       = {36},
  number       = {4},
  pages        = {744--755},
  year         = {2014},
  url          = {https://doi.org/10.1109/TPAMI.2013.240},
  doi          = {10.1109/TPAMI.2013.240}
}

@article{chen2024softmax,
  title={On softmax direct preference optimization for recommendation},
  author={Chen, Yuxin and Tan, Junfei and Zhang, An and Yang, Zhengyi and Sheng, Leheng and Zhang, Enzhi and Wang, Xiang and Chua, Tat-Seng},
  journal={Advances in Neural Information Processing Systems},
  volume={37},
  pages={27463--27489},
  year={2024}
}

@inproceedings{song2024preference,
  title={Preference ranking optimization for human alignment},
  author={Song, Feifan and Yu, Bowen and Li, Minghao and Yu, Haiyang and Huang, Fei and Li, Yongbin and Wang, Houfeng},
  booktitle={Proceedings of the AAAI Conference on Artificial Intelligence},
  volume={38},
  number={17},
  pages={18990--18998},
  year={2024}
}

@inproceedings{zhang2026value,
  title={Value: Value-aware large language model for query rewriting via weighted trie in sponsored search},
  author={Zhang, Xiao and Chen, Guanyu and Zuo, Boyang and Li, Feng and Wang, Pengjie and Xu, Jian and Zheng, Bo},
  booktitle={Proceedings of the 32nd ACM SIGKDD Conference on Knowledge Discovery and Data Mining V. 1},
  pages={2584--2593},
  year={2026}
}

@article{rafailov2023direct,
  title={Direct preference optimization: Your language model is secretly a reward model},
  author={Rafailov, Rafael and Sharma, Archit and Mitchell, Eric and Manning, Christopher D and Ermon, Stefano and Finn, Chelsea},
  journal={Advances in neural information processing systems},
  volume={36},
  pages={53728--53741},
  year={2023}
}

@misc{chidambaram2026robust,
  author = {Chidambaram, Keertana and Krishnamurthy, Sanath Kumar and Xu, Qiuling and Hsiao, Ko-Jen and Bhattacharya, Moumita},
  title = {Robust Post-Training for Generative Recommenders: Why Exponential Reward-Weighted {SFT} Outperforms {RLHF}},
  year = {2026},
  eprint = {2603.10279},
  archivePrefix = {arXiv},
  primaryClass = {cs.LG},
  doi = {10.48550/arXiv.2603.10279}
}

@inproceedings{zhou2023genrrl,
  author = {Zhou, Yujia and Dou, Zhicheng and Wen, Ji-Rong},
  title = {Enhancing Generative Retrieval with Reinforcement Learning from Relevance Feedback},
  booktitle = {Proceedings of the 2023 Conference on Empirical Methods in Natural Language Processing},
  year = {2023},
  month = dec,
  pages = {12481--12490},
  publisher = {Association for Computational Linguistics},
  address = {Singapore},
  doi = {10.18653/v1/2023.emnlp-main.768},
  url = {https://aclanthology.org/2023.emnlp-main.768/}
}

@inproceedings{zhang2025gpr,
  author    = {Zhang, Jun and Li, Yi and Liu, Yue and Wang, Changping and Wang, Yuan and
               Xiong, Yuling and Liu, Xun and Wu, Haiyang and Li, Qian and Zhang, Enming and
               Sun, Jiawei and Xu, Xin and Zhang, Zishuai and Liu, Ruoran and Huang, Suyuan and
               Zhang, Zhaoxin and Guo, Zhengkai and Yang, Shuojin and Guo, Meng-Hao and
               Yu, Huan and Jiang, Jie and Hu, Shi-Min},
  title     = {{GPR}: Towards a Generative Pre-trained One-Model Paradigm for Large-Scale Advertising Recommendation},
  booktitle = {Proceedings of the 32nd ACM SIGKDD Conference on Knowledge Discovery and Data Mining},
  year      = {2026},
  publisher = {ACM},
  doi       = {10.1145/3770855.3818425},
  url       = {https://doi.org/10.1145/3770855.3818425}
}

@inproceedings{2022dsi,
  author = {Tay, Yi and Tran, Vinh and Dehghani, Mostafa and Ni, Jianmo and Bahri, Dara and Mehta, Harsh and Qin, Zhen and Hui, Kai and Zhao, Zhe and Gupta, Jai and Schuster, Tal and Cohen, William W. and Metzler, Donald},
  title = {Transformer Memory as a Differentiable Search Index},
  booktitle = {Advances in Neural Information Processing Systems},
  volume = {35},
  pages = {21831--21843},
  publisher = {Curran Associates},
  year = {2022},
  doi = {10.52202/068431-1587},
  url = {https://proceedings.neurips.cc/paper_files/paper/2022/hash/892840a6123b5ec99ebaab8be1530fba-Abstract-Conference.html}
}

@inproceedings{nci,
  author = {Wang, Yujing and Hou, Yingyan and Wang, Haonan and Miao, Ziming and Wu, Shibin and Chen, Qi and Xia, Yuqing and Chi, Chengmin and Zhao, Guoshuai and Liu, Zheng and Xie, Xing and Sun, Hao and Deng, Weiwei and Zhang, Qi and Yang, Mao},
  title = {A Neural Corpus Indexer for Document Retrieval},
  booktitle = {Advances in Neural Information Processing Systems},
  volume = {35},
  pages = {25600--25614},
  publisher = {Curran Associates, Inc.},
  year = {2022},
  doi = {10.52202/068431-1856},
  url = {https://proceedings.neurips.cc/paper_files/paper/2022/hash/a46156bd3579c3b268108ea6aca71d13-Abstract-Conference.html}
}

@inproceedings{seal,
  author = {Bevilacqua, Michele and Ottaviano, Giuseppe and Lewis, Patrick and Yih, Scott and Riedel, Sebastian and Petroni, Fabio},
  title = {Autoregressive Search Engines: Generating Substrings as Document Identifiers},
  booktitle = {Advances in Neural Information Processing Systems},
  volume = {35},
  pages = {31668--31683},
  publisher = {Neural Information Processing Systems Foundation, Inc.},
  year = {2022},
  doi = {10.52202/068431-2296},
  url = {https://doi.org/10.52202/068431-2296}
}

@inproceedings{p5,
  author = {Geng, Shijie and Liu, Shuchang and Fu, Zuohui and Ge, Yingqiang and Zhang, Yongfeng},
  title = {Recommendation as Language Processing ({RLP}): A Unified Pretrain, Personalized Prompt \& Predict Paradigm ({P5})},
  booktitle = {Proceedings of the 16th ACM Conference on Recommender Systems},
  pages = {299--315},
  publisher = {ACM},
  year = {2022},
  doi = {10.1145/3523227.3546767},
  url = {https://doi.org/10.1145/3523227.3546767}
}

@inproceedings{2023tiger,
  author = {Rajput, Shashank and Mehta, Nikhil and Singh, Anima and Hulikal Keshavan, Raghunandan and Vu, Trung and Heldt, Lukasz and Hong, Lichan and Tay, Yi and Tran, Vinh and Samost, Jonah and Kula, Maciej and Chi, Ed and Sathiamoorthy, Maheswaran},
  title = {Recommender Systems with Generative Retrieval},
  booktitle = {Advances in Neural Information Processing Systems},
  volume = {36},
  pages = {10299--10315},
  publisher = {Curran Associates, Inc.},
  year = {2023},
  doi = {10.52202/075280-0452},
  url = {https://proceedings.neurips.cc/paper_files/paper/2023/hash/20dcab0f14046a5c6b02b61da9f13229-Abstract-Conference.html}
}

@inproceedings{2024lcrec,
  author = {Zheng, Bowen and Hou, Yupeng and Lu, Hongyu and Chen, Yu and Zhao, Wayne Xin and Chen, Ming and Wen, Ji-Rong},
  title = {Adapting Large Language Models by Integrating Collaborative Semantics for Recommendation},
  booktitle = {2024 IEEE 40th International Conference on Data Engineering (ICDE)},
  pages = {1435--1448},
  publisher = {IEEE},
  year = {2024},
  doi = {10.1109/ICDE60146.2024.00118},
  url = {https://doi.org/10.1109/ICDE60146.2024.00118}
}

@inproceedings{letter,
  author = {Wang, Wenjie and Bao, Honghui and Lin, Xinyu and Zhang, Jizhi and Li, Yongqi and Feng, Fuli and Ng, See-Kiong and Chua, Tat-Seng},
  title = {Learnable Item Tokenization for Generative Recommendation},
  booktitle = {Proceedings of the 33rd ACM International Conference on Information and Knowledge Management},
  pages = {2400--2409},
  publisher = {ACM},
  year = {2024},
  doi = {10.1145/3627673.3679569},
  url = {https://doi.org/10.1145/3627673.3679569}
}

@inproceedings{etegrec,
  author = {Liu, Enze and Zheng, Bowen and Ling, Cheng and Hu, Lantao and Li, Han and Zhao, Wayne Xin},
  title = {Generative Recommender with End-to-End Learnable Item Tokenization},
  booktitle = {Proceedings of the 48th International ACM SIGIR Conference on Research and Development in Information Retrieval},
  pages = {729--739},
  publisher = {ACM},
  year = {2025},
  doi = {10.1145/3726302.3729989},
  url = {https://doi.org/10.1145/3726302.3729989}
}

@inproceedings{diger,
  author = {Fu, Junchen and Ge, Xuri and Karatzoglou, Alexandros and Arapakis, Ioannis and Verberne, Suzan and Jose, Joemon M. and Ren, Zhaochun},
  title = {Differentiable Semantic {ID} for Generative Recommendation},
  booktitle = {Proceedings of the 49th International ACM SIGIR Conference on Research and Development in Information Retrieval},
  pages = {369--379},
  publisher = {ACM},
  year = {2026},
  doi = {10.1145/3805712.3809641},
  url = {https://doi.org/10.1145/3805712.3809641}
}

@misc{2025onerec,
  author = {Deng, Jiaxin and Wang, Shiyao and Cai, Kuo and Ren, Lejian and Hu, Qigen and Ding, Weifeng and Luo, Qiang and Zhou, Guorui},
  title = {{OneRec}: Unifying Retrieve and Rank with Generative Recommender and Iterative Preference Alignment},
  year = {2025},
  eprint = {2502.18965},
  archivePrefix = {arXiv},
  doi = {10.48550/arXiv.2502.18965},
}

@misc{chen2025onesearch,
  author = {Chen, Ben and Guo, Xian and Wang, Siyuan and Liang, Zihan and Lv, Yue and Ma, Yufei and Xiao, Xinlong and Xue, Bowen and Zhang, Xuxin and Yang, Ying and Dai, Huangyu and Xu, Xing and Zhao, Tong and Peng, Mingcan and Zheng, Xiaoyang and Wang, Chao and Zhao, Qihang and Zhai, Zhixin and Zhao, Yang and Liu, Bochao and Lv, Jingshan and Liang, Xiao and Ding, Yuqing and Chen, Jing and Lei, Chenyi and Ou, Wenwu and Li, Han and Gai, Kun},
  title = {{OneSearch}: A Preliminary Exploration of the Unified End-to-End Generative Framework for E-commerce Search},
  year = {2025},
  eprint = {2509.03236},
  archivePrefix = {arXiv},
  doi = {10.48550/arXiv.2509.03236},
  url = {https://arxiv.org/abs/2509.03236v5}
}

@misc{wan2026r3,
  author = {Wan, Qiang and Yang, Ze and Yang, Dawei and Fan, Ying and Yan, Xin and Liu, Siyang and Liu, Yicong and Zhang, Chenwei and Xu, Wei and Qin, Jiahao and Wang, Ke},
  title = {{R3-VAE}: Reference Vector-Guided Rating Residual Quantization {VAE} for Generative Recommendation},
  year = {2026},
  eprint = {2604.11440},
  archivePrefix = {arXiv},
  doi = {10.48550/arXiv.2604.11440},
  url = {https://arxiv.org/abs/2604.11440v3}
}

@misc{varg,
  author = {Xiaopeng Chu and Jianbo Zhu and Mingmin Jin and Jing Wang and Xing Fang and Wenyi Zhang},
  title = {{VARG}: Value-Aware and Ranking-Aligned Generative Retrieval for Dynamic E-commerce Search},
  year = {2026},
  eprint = {2609.14493},
  archivePrefix = {arXiv},
  url = {https://arxiv.org/abs/2609.14493}
}

@misc{fu2026rankgr,
  author = {Fu, Kairui and Wu, Changfa and Yuan, Kun and Cao, Binbin and Huang, Dunxian and Yan, Yuliang and Zheng, Junjun and Zhang, Jianning and Zhou, Silu and Wu, Jian and Kuang, Kun},
  title = {{RankGR}: Rank-Enhanced Generative Retrieval with Listwise Direct Preference Optimization in Recommendation},
  year = {2026},
  eprint = {2602.08575},
  archivePrefix = {arXiv},
  doi = {10.48550/arXiv.2602.08575},
  url = {https://arxiv.org/abs/2602.08575}
}

@inproceedings{raddpo,
  author = {Chen, Zhiguo and Sun, Guohao and Qiu, Yiming and Yao, Xingzhi and Li, Mingming and Wang, Huimu and Zhang, Yangqi and Wang, Songlin and Xu, Sulong},
  title = {{RAD-DPO}: Robust Adaptive Denoising Direct Preference Optimization for Generative Retrieval in E-commerce},
  booktitle = {Proceedings of the 49th International ACM SIGIR Conference on Research and Development in Information Retrieval},
  pages = {4507--4512},
  publisher = {ACM},
  year = {2026},
  doi = {10.1145/3805712.3808512},
  url = {https://doi.org/10.1145/3805712.3808512}
}

@misc{schulman2017proximal,
  author = {Schulman, John and Wolski, Filip and Dhariwal, Prafulla and Radford, Alec and Klimov, Oleg},
  title = {Proximal Policy Optimization Algorithms},
  year = {2017},
  eprint = {1707.06347},
  archivePrefix = {arXiv},
  doi = {10.48550/arXiv.1707.06347},
  url = {https://arxiv.org/abs/1707.06347}
}

@misc{shao2024deepseekmath,
  author = {Shao, Zhihong and Wang, Peiyi and Zhu, Qihao and Xu, Runxin and Song, Junxiao and Bi, Xiao and Zhang, Haowei and Zhang, Mingchuan and Li, Y. K. and Wu, Y. and Guo, Daya},
  title = {{DeepSeekMath}: Pushing the Limits of Mathematical Reasoning in Open Language Models},
  year = {2024},
  eprint = {2402.03300},
  archivePrefix = {arXiv},
  doi = {10.48550/arXiv.2402.03300},
  url = {https://arxiv.org/abs/2402.03300v3}
}

@misc{zhang2026hcgrec,
  author = {Zhang, Kangning and Fang, Haotian and Luo, Xukun and Yin, Hao and Gao, Yang and Yan, Peng and Liu, Weiwen and Zhang, Weinan and Yu, Yong},
  title = {Learning from Unreachable Rewards: Hint-Conditioned Reinforcement Learning for Generative Recommendation},
  year = {2026},
  eprint = {2608.11980},
  archivePrefix = {arXiv},
  doi = {10.1145/3799682.3840842},
  url = {https://arxiv.org/abs/2608.11980v2}
}

@misc{chen2026onesearch,
  author = {Chen, Ben and Wang, Siyuan and Ma, Yufei and Liang, Zihan and Zhang, Xuxin and Lv, Yue and Yang, Ying and Dai, Huangyu and Mao, Lingtao and Zhao, Tong and Qian, Zhipeng and Sun, Xinyu and Zhai, Zhixin and Zhao, Yang and Liu, Bochao and Lv, Jingshan and Liang, Xiao and Kong, Hui and Chen, Jing and Li, Han and Lei, Chenyi and Ou, Wenwu and Gai, Kun},
  title = {{OneSearch-V2}: The Latent Reasoning Enhanced Self-distillation Generative Search Framework},
  year = {2026},
  eprint = {2603.24422},
  archivePrefix = {arXiv},
  doi = {10.48550/arXiv.2603.24422},
  url = {https://arxiv.org/abs/2603.24422v2}
}

@inproceedings{hrpo,
  author = {Guo, Kaifeng and Yang, Yiming and Gao, Jingtong and Zeng, Guolei and Yang, Fukang and Liang, Yukang and Jiang, Peng and Cai, Qingpeng and Zhao, Xiangyu},
  title = {Hierarchical Residual Policy Optimization for Generative Recommendations},
  booktitle = {Proceedings of the 32nd ACM SIGKDD Conference on Knowledge Discovery and Data Mining V.2},
  pages = {1427--1438},
  publisher = {ACM},
  year = {2026},
  doi = {10.1145/3770855.3818206},
  url = {https://doi.org/10.1145/3770855.3818206}
}

@misc{chen2026localsug,
  author = {Chen, Jinwen and Zhang, Shiwen and Gong, Shuai and Zhang, Zheng and Zhao, Yachao and Wang, Lingxiang and Zhou, Haibo and Lin, Wei and Zhang, Hainan},
  title = {{LocalSUG}: City-Preference-Enhanced {LLM} for Query Suggestion in Local-Life Services},
  year = {2026},
  eprint = {2603.04946},
  archivePrefix = {arXiv},
  doi = {10.48550/arXiv.2603.04946},
  url = {https://arxiv.org/abs/2603.04946v3}
}

@misc{dhariwal2020jukebox,
  author = {Dhariwal, Prafulla and Jun, Heewoo and Payne, Christine and Kim, Jong Wook and Radford, Alec and Sutskever, Ilya},
  title = {{Jukebox}: A Generative Model for Music},
  year = {2020},
  eprint = {2005.00341},
  archivePrefix = {arXiv},
  doi = {10.48550/arXiv.2005.00341},
  url = {https://arxiv.org/abs/2005.00341}
}

@misc{zhang2026onemall,
  author = {Zhang, Kun and Zhang, Jingming and Cheng, Wei and Cheng, Yansong and Zhang, Jiaqi and Lu, Hao and Zhang, Xu and Gan, Haixiang and Cao, Jiangxia and Wang, Tenglong and Zhang, Ximing and Xia, Boyang and Cai, Kuo and Wang, Shiyao and Dou, Hongjian and Yu, Jinkai and Wen, Mingxing and Luo, Qiang and Liang, Dongxu and Lei, Chenyi and Wang, Jun and Liu, Runan and Liu, Zhaojie and Tang, Ruiming and Gao, Tingting and Liu, Shaoguo and Ding, Yuqing and Kong, Hui and Li, Han and Zhou, Guorui and Ou, Wenwu and Gai, Kun},
  title = {{OneMall}: One Architecture, More Scenarios --- End-to-End Generative Recommender Family at Kuaishou E-Commerce},
  year = {2026},
  eprint = {2601.21770},
  archivePrefix = {arXiv},
  doi = {10.48550/arXiv.2601.21770},
  url = {https://arxiv.org/abs/2601.21770v2}
}

@misc{2024GenR-PO, 
  title={Generative Retrieval with Preference Optimization for E-commerce Search}, author={Mingming Li and Huimu Wang and Zuxu Chen and Guangtao Nie and Yiming Qiu and et al.}, 
  year={2024}, 
  eprint={2407.19829}, 
  archivePrefix={arXiv}, 
  primaryClass={cs.IR}, 
  url={https://arxiv.org/abs/2407.19829}
}

@article{xia2026unleash,
  title={Unleash the potential of long semantic ids for generative recommendation},
  author={Xia, Ming and Ma, Guoxin and Zhou, Zhiqin and Huang, Dongmin},
  journal={arXiv preprint arXiv:2602.13573},
  year={2026}
}

@article{radford2019gpt2,
  title={Language models are unsupervised multitask learners},
  author={Radford, Alec and Wu, Jeffrey and Child, Rewon and Luan, David and Amodei, Dario and Sutskever, Ilya and others},
  journal={OpenAI blog},
  volume={1},
  number={8},
  pages={9},
  year={2019}
}

@article{oord2018representation,
  title={Representation learning with contrastive predictive coding},
  author={Oord, Aaron van den and Li, Yazhe and Vinyals, Oriol},
  journal={arXiv preprint arXiv:1807.03748},
  year={2018}
}

\end{document}